\documentclass[prb,aps,twocolumn,10pt]{revtex4-1}
\usepackage[utf8x]{inputenc}
\usepackage{amsmath,amssymb,graphicx,xcolor,esint,bm}

\newcommand{\re}[1]{\mbox{Re}\left\{#1\right\}}

\newcommand{\im}[1]{\mbox{Im}\left\{#1\right\}}

\newif\ifpaper
\papertrue

\begin{document}
\title{Electromagnetic Scattering Resonances \\ of Quasi-1D Nanoribbons}
\author{Carlo Forestiere}
\affiliation{ Department of Electrical Engineering and Information Technology, Universit\`{a} degli Studi di Napoli Federico II, via Claudio 21,
 Napoli, 80125, Italy}
\author{Giovanni Miano}
\affiliation{ Department of Electrical Engineering and Information Technology, Universit\`{a} degli Studi di Napoli Federico II, via Claudio 21,
 Napoli, 80125, Italy}
 \author{Mariano Pascale}
\affiliation{ Department of Electrical Engineering and Information Technology, Universit\`{a} degli Studi di Napoli Federico II, via Claudio 21,
 Napoli, 80125, Italy}
\author{Roberto Tricarico}
\affiliation{ Department of Electrical Engineering and Information Technology, Universit\`{a} degli Studi di Napoli Federico II, via Claudio 21,
 Napoli, 80125, Italy}

%\section{Formulation}
%\
\begin{abstract}
We analyse the resonance conditions of a long and narrow ribbon of finite length whether it is conductive or dielectric. This is accomplished by using a full wave approach based on the material independent modes that naturally discriminates the role of the geometry and of the material. This method effectively allows the design of the material in such a way to obtain the desired resonances. Eventually, as an example, we design two quasi-one dimensional  resonators based on a graphene layer and on a silicon thin film.
\end{abstract}
\maketitle

\section{Introduction}
Downscaling electromagnetic resonators  remains a major issue in micro and nanotechnology and asks for novel platforms supporting electromagnetic waves and resonances on lower dimensional structures. 

In the last years, the emerging spatial localization properties of plasmons in noble metal structures have suggested several strategies to squeeze the electromagnetic energy over subwavelength spatial regions by using, for instance, a coated metal pin \cite{takahara1997guiding}, V-shaped grooves  \cite{bozhevolnyi2006channel} and wedges \cite{moreno2008guiding}. However, in noble metals, plasmons have intrinsic limitations \cite{khurgin2015deal}, including, above all, the  short lifetime due to the metal losses and the limited spectral tunability, spanning only from the ultraviolet to the near infrared. 

Electromagnetic waveguiding in lower dimensional spatial domains has been demonstrated in photonic crystals (PCs), where a careful design of the reciprocal space enables the creation of interfaces supporting topologically protected one-directional propagation \cite{lu2014topological}. Unfortunately, PCs necessarily require the fabrication of large structures. Properly engineered metasurfaces can also play a central role in conceiving electromagnetic circuitry of reduced dimensionality. Recently, Bisharat et al. have demonstrated that two planar surfaces of complementary surface impedance may guide electromagnetic waves along a one dimensional line \cite{PhysRevLett.119.106802}. However, a resonator based on these  modes would require the use of semi-infinite metasurfaces (or at least very large compared to the operating wavelength) with a defect-free interface. These two characteristics make it bulky and difficult to downscale.

Another promising platform is represented by two dimensional materials for example doped graphene \cite{grigorenko2012graphene} and transition metal dichalcogenides \cite{krasnok2018nanophotonics}.
 Specifically, thanks to the high carrier mobility,  graphene plasmons can have a lifetime that may reach  hundreds of optical cycles, one order of magnitude greater than noble metal ones.  Graphene ribbons of infinite length and finite width have been proposed as waveguides and the  properties of their modes have been investigated \cite{christensen2011graphene,nikitin2011edge}. 
The corresponding mode patterns  have been experimentally observed \cite{Fei:15}.

A long and narrow ribbon of finite length,  whether it is conductive or dielectric, may  behave  as a quasi-one dimensional (1D) electromagnetic resonator. %The word quasi one - dimensional is a reminder that the structure is, strictly speaking, two - dimensional. Nevertheless, the quantities of physical interest are described, with excellent approximation, by one - dimensional integro - differential equations. 
For an effective analysis and design it would be highly desirable to know its resonances and its resonant modes and how they depend on its geometrical and physical parameters. The material independent modes \cite{Forestiere16} provide a unified approach that allows to  clearly separate the roles of the material and of the geometry. 

In this paper, by using a full wave approach based on the material independent modes, we derive, for the first time, the resonance conditions of a quasi 1D structure of finite length whether it is conducting or dielectric. In particular, these conditions enable us to determine the relations between the material parameters, the geometrical parameters, and the wavelength so that the ribbon resonates.  We also show that, the electric field at the ends of the ribbon undergoes a strong enhancement due to the strong charge accumulation at the ends. This property is a consequence of the structure of the integro-differential operator characterizing the problem, which does not depend on the particular material.
 
The paper is organized as follows. In Sec. \ref{sec:Formulation} we formulate the electromagnetic scattering problem for a long and narrow ribbon of finite length. Under this hypothesis, the problem reduces with excellent approximation to the solution of a 1D integro-differential equation, which is solved by using the material-
independent modes. In Sec.  \ref{sec:Results}, we analyse the modes of the ribbon, considering both the cases of a length-to-wavelength ratio much smaller than one, and comparable to one. In Sec. \ref{sec:Virtual}, we consider the scattering problem from a ribbon with tunable conductivity, and we calculate the electric field distribution and the scattering efficiency under a plane-wave excitation. 
Eventually, as an example,  we design  in Sec. \ref{sec:Design} the resonant scattering from a doped graphene ribbon and from a silicon film in the infrared spectral range. In both cases we examine the enhancement and the localization properties of the electric field.

\section{Electromagnetic scattering from a ribbon}
\label{sec:Formulation}
\begin{figure}
\centering
\includegraphics[width=\columnwidth]{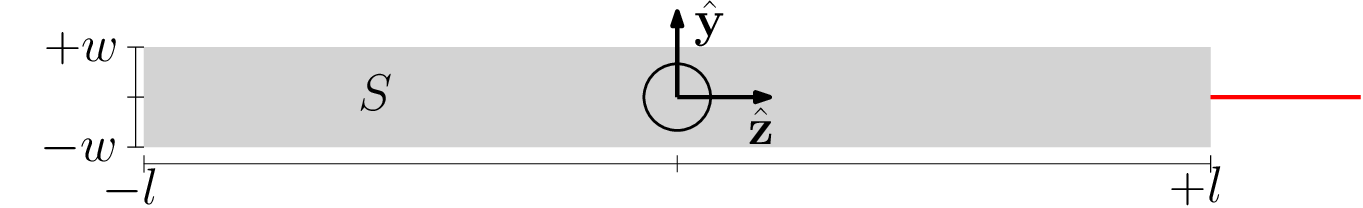}
\caption{Sketch of the ribbon of half-length $l$ and half-width $w$.}
  \label{fig:Sketch}
\end{figure}

We consider a body with thickness $\Delta$ much smaller than its other two linear characteristic dimensions and than the wavelength of the electromagnetic field. In this limit only the in-plane electromagnetic response of the material is important and the body may be treated as it is two-dimensional (2D) \cite{Harrington}.  The 2D homogeneous material has a rectangular shape as sketched in Fig. \ref{fig:Sketch}. The rectangle has  length $2l$, width $2w$, and it is aligned along the $\hat{\bf z}$ axis of a Cartesian coordinate system. It is illuminated by an electric field with angular frequency $\omega$, i.e. ${\bf e}_i  \left( {\bf r}, t \right) = \re{{\bf E}_i \left( {\bf r}\right) e^{i \omega t}}$. We assume that the rectangle is a ribbon with very high length-to-width ratio $l/w \gg 1$ and $k_0 w \ll 1$,  where $k_0 = \omega/c$ and $c$ is the speed of light in vacuum. 	

\subsection{Integro-differential Equation for the Induced Current}
In the following, we introduce the quasi-1D model for the induced current along the conducting or dielectric ribbon. In both cases the sheet is characterized by an effective surface conductivity $\sigma \left( \omega \right)$.  In the linear regime, the electric field induces a surface free or polarization current density ${\bf j}_S \left( {\bf r}_s, t \right) = \re{{\bf J} \left( {\bf r}_s\right) e^{i \omega t}}$ and a surface free or polarization charge density ${\bf \rho}_S \left( {\bf r}_s, t \right) = \re{{ \Sigma} \left( {\bf r}_s\right) e^{i \omega t}}$, with ${\bf r}_s \in S$, where $S$ is the surface of the ribbon. We indicate with ${\bf e} \left( {\bf r}_s, t \right) = \re{{\bf E} \left( {\bf r}_s \right) e^{i \omega t}}$ the total electric field, i.e. the sum of the impressed and the scattered electric fields. Since $l/w \gg 1$ and $k_0 w \ll 1$, we  disregard the spatial variation along the $y$ direction of the various physical quantities. Furthermore, we consider only transverse magnetic (TM) excitation, and we also disregard the transverse component of the current density field. 
 Therefore, we set ${\bf J} \left( \mathbf{r}_s \right)  =  I \left( z \right)/2w \, \hat{\bf z}$, $\Sigma \left( \mathbf{r}_s \right)  = Q \left( z \right)/2w$,  and $ {\bf E} \left( \mathbf{r}_s \right) = {E} \left( z \right) \, \hat{\bf z}$,   where $I \left( z \right)$ is the current intensity through the ribbon transverse section,  $Q\left( z \right)$ is the per unit length (p.u.l.) charge, and $-l < z < l$. In general, for a TM excitation, the longitudinal current density behaves as $ I \left( z \right)/\left( \pi  \sqrt{w^2-y^2} \right)$ when $y \rightarrow \pm w $ \cite{Butler80}. By assuming the  uniformity of the current density along $\bf{\hat y}$,  we make a negligible error as long as $l/w \gg 1$ and $k_0 w \ll 1$. This is shown in Sec. \ref{sec:Results}, where we compare the solution obtained by the quasi-1D model with the one obtained by a fully 2D simulation.
 
 In the frequency domain, the current intensity on the ribbon is governed by the constitutive equation
\begin{equation}
 E \left( z \right) =
\frac{1}{2 w \sigma \left( \omega \right)}  I \left( z \right)  \qquad \mbox{for} \, \left| z \right| < l.
  \label{eq:SurfCond} 
\end{equation}
In the following, we disregard the effects of the spatial dispersion. On the other hand, the axial component of the total electric field on the ribbon surface is given by
\begin{equation}
   E \left( z \right) = - i \omega A - \frac{dV}{dz} + E_i \qquad \mbox{for} \, \left| z \right| < l,
   \label{eq:Erepresentation}
\end{equation}
where $A \left( z \right)$ and $ V \left( z \right)$ are, respectively, the axial components of the induced magnetic vector potential and of the induced electric scalar potential, evaluated on the ribbon axis and $E_i \left( z \right) = \hat{\bf z} \cdot {\bf E}_i \left( x= 0,y=0,z\right)$. By using the Lorenz gauge we obtain
\begin{equation}
\begin{aligned}
  A \left( z \right) &= \mu_0 \, \mathcal{L} \left\{ I \right\} \left( z \right) \qquad \mbox{for} \, \left| z \right| < l, \\
  V \left( z \right) &= \frac{1}{\varepsilon_0} \, \mathcal{L} \left\{ Q \right\} \left( z \right)  \qquad \mbox{for} \, \left| z \right| < l,
\end{aligned}
\label{eq:Potentials}
\end{equation}
where $\mathcal{L} \left\{ u \right\}$ is the linear integral operator
\begin{equation}
   \mathcal{L} \left\{ u \right\} \left( z \right) = \frac{1}{4 \pi} \int_{-l}^{+l} g \left( z - z' \right) u \left( z ' \right) dz' \quad \mbox{for} \, \left| z \right| < l,
\end{equation}
\begin{equation}
  g \left( \zeta \right) = \frac{1}{2w} \int_{-w}^{+w} {r}^{-1} \left( y, \zeta \right) \exp \left[-ik_0 {r} \left( y, \zeta \right)\right] dy,
  \end{equation}
and 
\begin{equation}
{ r} = \sqrt{y^2+\zeta^2}.
\end{equation}
It is convenient to express the Green function $g$ as $g = g_S + g_D$,  where $g_S$ is the static Green function
\begin{equation}
g_S \left( \zeta \right) = \frac{1}{2w} \ln\left[ \frac{+1+\sqrt{1+\left(\zeta/w\right)^2}}{-1+\sqrt{1+\left(\zeta/w\right)^2}} \right].
\label{eq:StaticGreen}
\end{equation}
Since $k_0 w \ll 1 $ the dynamic contribution $g_D$ is approximated as, e.g. \cite{Franceschetti}
\begin{equation}
g_D \left( \zeta \right) \approx -ik_0 \, \mbox{sinc} \left( k_0 \left| \zeta \right| / 2 \right) \exp \left( - ik_0 \left| \zeta \right| \right).
\label{eq:DynamicGreen}
\end{equation}
By combining Eqs. (\ref{eq:SurfCond}),(\ref{eq:Erepresentation}),(\ref{eq:Potentials}) and the continuity equation
\begin{equation}
   \frac{dI}{dz} = -i \omega Q,
\end{equation}
 we obtain the equation for the distribution of the current intensity along the ribbon axis
\begin{equation}
\Gamma I - \mathcal{F} \left\{ I \right\} \left( z \right) = i 2 w  E_i \left( z \right) \qquad \mbox{for} \, \left| z \right| < l,
\label{eq:IntegroDifferential}
\end{equation}
where
\begin{equation}
\Gamma = \frac{i}{\sigma},
\label{eq:GammaDef}
\end{equation}
 $\mathcal{F}$ takes into account the induced electric field
\begin{equation}
 \mathcal{F} \left\{ I \right\} \left( z \right) = \frac{2 w \zeta_0}{k_0} \left[ \frac{d}{dz} \mathcal{L} \left\{ \frac{dI}{dz} \right\} \left( z \right) + k_0^2 \mathcal{L} \left\{ {I} \right\} \left( z \right) \right],
 \label{eq:OperatorF}
\end{equation}
and $\zeta_0 = \sqrt{\mu_0/\varepsilon_0}$.
The integro-differential equation (\ref{eq:IntegroDifferential}) has to be solved with the boundary conditions
\begin{equation}
 I \left( -l \right) = I \left( + l \right) = 0.
 \label{eq:BC}
\end{equation}
The same equation also holds for a conducting tubule, assuming as static Green function  
\begin{equation}
   g_S \left( \zeta \right) =\frac{2}{\pi} \frac{K \left( m \right)}{\sqrt{4a^2 + \zeta^2}},
\end{equation}
where $K \left( m \right)$ is the complete elliptic integral of the first kind,
\begin{equation}
 m = \frac{4 a^2}{4a^2+\zeta^2},
\end{equation}
and $a$ is the tubule radius.
\subsection{Solution in terms of material independent modes}
We solve Eq. (\ref{eq:IntegroDifferential}) with the boundary conditions (\ref{eq:BC}) by using the material-independent modes \cite{Forestiere16}. They are the solution of the eigenvalue problem
\begin{equation}
   \mathcal{F} \left\{ u \right\} \left( z \right) = \gamma \, u \left( z \right) \qquad \mbox{for} \, \left| z \right| < l,
   \label{eq:EigenvalueProblem}
\end{equation}
with the boundary condition \ref{eq:BC}, where $u\left( z \right)$  is the eigenfunction associated with the eigenvalue $\gamma$. As for 3D objects, the operator $\mathcal{F}$ is compact, thus its spectrum $\left\{\gamma_n \right\}_{n \in \mathbb{N}}$ is countable infinite, but $\mathcal{F}$ is not Hermitian because of the radiation losses \cite{Lalanne,forestiere2018volume}. The eigenvalues are complex. We have
\begin{equation}
\begin{aligned}
 \mbox{Re} \left\{ \gamma_n \right\} &=   \frac{ 8 w \omega }{\int_{-l}^{+l} \left| u_n \right|^2 dz} \times \; \\ &\times \left( \frac{\mu_0}{4}  \iint_{\mathbb{R}^3} \left\| \mathbf{H}_n \right\|^2 dV - \frac{\varepsilon_0}{4}  \iint_{\mathbb{R}^3} \left\| \mathbf{E}_n \right\|^2 dV \right), \\
 \mbox{Im} \left\{ \gamma_n \right\} &=  - \frac{ 4 w }{\int_{-l}^{+l} \left| u_n \right|^2 dz}  \; \frac{1}{2 \zeta_0} \oiint_{S_\infty} \left\|\mathbf{E}_n \right\|^2 \, dS,
 \end{aligned}
 \label{eq:Poynting}
\end{equation}
where $\mathbf{E}_n$ and $\mathbf{H}_n$ are the electric and magnetic fields radiated by the current  $u_n$, and $S_\infty$ is a spherical surface with infinite radius. The real part of the eigenvalue is proportional to the difference between the magnetic and the electric energies of the mode. Thus, it is negative when the electric energy is greater than the magnetic one, positive otherwise. The imaginary part of the eigenvalue is negative and it is proportional to the power radiated to infinity by the corresponding mode, therefore it takes into account the radiation losses.

The eigenmodes $u_n \left( z \right)$ and $u_m \left( z \right)$, corresponding to two different eigenvalues, are not orthogonal in the usual sense, i.e. $\langle u_n, u_m\rangle \ne 0$, where $\langle u, v \rangle = \int u^* \left( z \right) v \left( z \right) dz$. Nevertheless, we have $\langle  u_n^*, u_m\rangle = 0$ for $n \ne m$. Moreover, due to the symmetry of the problem, the eigenmodes are either even or odd functions of $z$. In the limit $k_0 l \ll 1$  the operator $\mathcal{F}$ is Hermitian because the radiation losses are negligible, its eigenvalues are real and negative, while its eigenmodes are real and orthogonal in the usual sense.

The solution of equation (\ref{eq:IntegroDifferential}) with the boundary conditions (\ref{eq:BC}) is therefore 
\begin{equation}
  I \left( z \right) = i 2 w  \displaystyle\sum_{h=1}^\infty \frac{1}{\Gamma-\gamma_n} \frac{\langle u_n^*, E_i \rangle}{\langle u_n^*, u_n \rangle }u_n.
\label{eq:ModeExpansion}
\end{equation}
The eigenvalues $\gamma_n$ and the eigenmodes ${ u}_n$ are material independent, they only depend on the quantities $l/w$ and $l/\lambda$, where $\lambda = 2\pi/k_0$. The material only appears through $\Gamma$ in the factors $ 1/\left( \Gamma - \gamma_n \right)$. Equation (\ref{eq:ModeExpansion}) distinctly separates the role of the geometry from the role played by the material. For assigned values of material, geometry, frequency, and excitation the expression (\ref{eq:ModeExpansion}) is computationally disadvantageous compared to the direct numerical solution of Eq. (\ref{eq:IntegroDifferential}). However, when the scattered field has to be computed for many values of surface conductivity (as in Sec. \ref{sec:Virtual}) or different excitation conditions, and the geometry and the frequency are assigned, the computation of Eq. (\ref{eq:ModeExpansion}) is computationally advantageous compared to the direct solution. The main advantage of solution (\ref{eq:ModeExpansion}) is that it gives us directly the resonances and the coupling of the modes with the incident field.

For passive materials we have $\im{\Gamma} \ge 0$, thus the quantity  $\left| \Gamma - \gamma_n   \right|$ in Eq. (\ref{eq:ModeExpansion}) does not vanish because $\im{\gamma_{n}}<0$. Nonetheless, the amplitude of the $n$-th mode increases as the distance between $\Gamma$ and $\gamma_n$ is reduced. If we assign the material and the geometrical dimensions of the ribbon, the resonance condition in the usual ``frequency picture" for the $n$-th mode is 
\begin{equation}
   \left| \Gamma \left( \lambda \right) - \gamma_n  \left( {l}/{\lambda} \right) \right| = \underset{\lambda}{\mbox{Minimum}}.
   \label{eq:ResonanceCondition2}
\end{equation}
It is possible to introduce a complementary view, denoted as ``material picture", where the dimensions of the ribbon and the operating wavelength are assigned. In this case, the resonance condition for the $n$-th mode is 
\begin{equation}
   \left| \Gamma - \gamma_n \right| = \underset{\Gamma}{\mbox{Minimum}}.
   \label{eq:ResonanceCondition}
\end{equation}
The ``material picture'' is particularly relevant because the conductivity of 2D materials, e.g. graphene ribbons, can be either tuned chemically or by electrostatic gating, while the effective conductivity  of a dielectric thin film can be tuned by varying its thickness. 

%Eq. \ref{eq:ResonanceCondition} is a universal resonance condition because it applies for any ribbon of given ratios $l/w$ and $l/\lambda$ regardless of the material. 

A passive material satisfying Eq. (\ref{eq:ResonanceCondition}) has $\re{\Gamma} = \re{\gamma_n}$.  The imaginary part of $\sigma$ and hence the real part of $\Gamma$ may be either negative or positive depending on the material and on the frequency. In particular, below the frequency where interband transitions occur, the real part of $\Gamma$ is negative for conductive materials and positive for dielectric materials. Therefore, in conducting materials modes with $\re{\gamma_n}<0$ can be resonantly excited, while in dielectric materials modes with $\re{\gamma_n} > 0$ can be resonantly excited. In section \ref{sec:Results}, we show that we can design the resonances in both ways.

%For instance for a silicon dioxide substrate with $\varepsilon_S=3.9$ the eigenvalues are multiplied by $0.41$.}	

\subsection{Approximated approach for $l/w \rightarrow \infty$}
The integro-differential problem introduced so far can be solved analytically in the limit $l/w \rightarrow 0 $, e.g. \cite{Franceschetti}. Specifically, the static Green function $g_S$ of Eq. (\ref{eq:StaticGreen}) has a singularity of logarithmic type at $\zeta = 0$ which prevails over the dynamic contribution of $g_D$. 
When $w \rightarrow 0 $ the function $g_S$ behaves as a Dirac delta function with amplitude (ribbon slenderness)
\begin{equation}
   \Theta = \int_{-l}^{l} g_S \left( \zeta \right) d \zeta,
\end{equation}
and it turns out that
\begin{equation}
     \mathcal{F} \left\{ I \right\} \left( z \right) \approx \mathcal{F}^{\left( \text{A} \right)} \left\{ I \right\} \left( z \right) =  \frac{2 w \zeta_0}{k_0} \frac{\Theta}{4\pi} \left( \frac{d^2 I}{d z^2} + k_0^2 I \right),
   \label{eq:FthinApprox}
\end{equation}
where 
\begin{equation}
   \Theta \approx 2 \ln \left( \frac{2l}{w} \right).
   \label{eq:Theta}
\end{equation}  
%If the function $u \left( z \right)$ and its first derivative vary slowly on the length scale of the order of $a$	then approximation \ref{eq:FthinApprox} is reasonable.
 In the following we denote the quantity obtained in this approximation with the superscript $A$. The expression  of $\Theta$ for a tubule of radius $a$ is analogous to Eq. (\ref{eq:Theta}), providing that $w$ is replaced by $a$. 

The eigenvalues $ \gamma_n^{\left(\text{A}\right)}$ and the eigenmodes $u_n^{\left(\text{A}\right)}$ of $\mathcal{F}^{\left( \text{A} \right)} \left\{ I \right\}$ are given by

\begin{equation}
   \gamma_n^{\left(\text{A}\right)} = \frac{2 w \zeta_0}{k_0} \frac{\Theta}{4\pi} \left(  - \beta_n^2 + k_0^2\right) \qquad n=0,1,2,3,\ldots,
    \label{eq:EigAnl}
\end{equation}

\begin{equation}
 u_n^{\left(\text{A}\right)} \left(z\right) = 
 \begin{cases}
  \cos{\beta_n z} \qquad n=0,2,4,6,\ldots, \\
  \sin{\beta_n z} \qquad n=1,3,5,7,\ldots, 
 \end{cases}
 \label{eq:ModesAnl}
\end{equation}

\begin{equation}
   \beta_n = \frac{\pi}{2l} \left( 1 + n \right). 
    \label{eq:BetaAnl}
\end{equation}

The approximated operator $\mathcal{F}^{\left( \text{A} \right)} \left\{ I \right\}$ does not take into account the finite length of the ribbon, which comes into play  only through the boundary conditions, expressed by Eq. (\ref{eq:BetaAnl}). As a consequence, this approximation is not able to predict the charge accumulation at the ribbon end, as we will see in the following section.   We note that the eigenvalues $\gamma_n^{\left( \text{A} \right) }$ and the eigenmodes $u_n^{\left(\text{A}\right)}$ are real because we have disregarded $g_D$. Even if this approximation disregards the retardation effects, it  takes into account both the quasi-static electric and magnetic interactions far away from the ribbon ends. Specifically, the  electric and the magnetic interactions are described by the first and the second term in parenthesis in Eq. (\ref{eq:FthinApprox}) ( they correspond to the first and second term in parenthesis in the expression of the eigenvalue (\ref{eq:EigAnl})). 

When $ k_0 \ll \beta_n$ the electric contribution is dominant and the mode has a quasi-electrostatic character. It is characterized by a negative eigenvalue $\gamma_n^{\left( \text{A} \right)}$. This condition certainly occurs when $l/\lambda \ll 1$. For $k_0 \approx \beta_n$ both the magnetic and the electric contributions are important and the mode has an electromagnetic character.

For a uniform $E_i$, the solution of Eq. (\ref{eq:IntegroDifferential}) with the boundary conditions (\ref{eq:BC}) is
\begin{equation}
  I^{\left( \text{A} \right)}\left( z \right) = - \frac{4\pi}{\Theta} \frac{i k_0 E_i}{\beta^2 \zeta_0}\left( 1 - \frac{\cos{\beta z}}{\cos{\beta l}} \right),
\label{eq:TWAnonhomSD}
\end{equation}
\begin{equation}
   \beta = \sqrt{k_0^2 - \frac{4 \pi}{\Theta} \frac{k_0}{2 w } \frac{\Gamma}{\zeta_0} }.
\end{equation}
In the weak losses limit, $\left|   I^{\left( \text{A} \right)}\left( z \right) \right|$ is maximum when $\re{\beta l} = \left(2n+1\right)\pi/2$ for $n=0,1,\ldots $. This condition is equivalent to either condition (\ref{eq:ResonanceCondition2}) or condition (\ref{eq:ResonanceCondition}) for the even modes (odd modes are not excited in this condition). 

\section{Modal Analysis}
\label{sec:Results}
\begin{figure}
\centering
\includegraphics[width=\columnwidth]{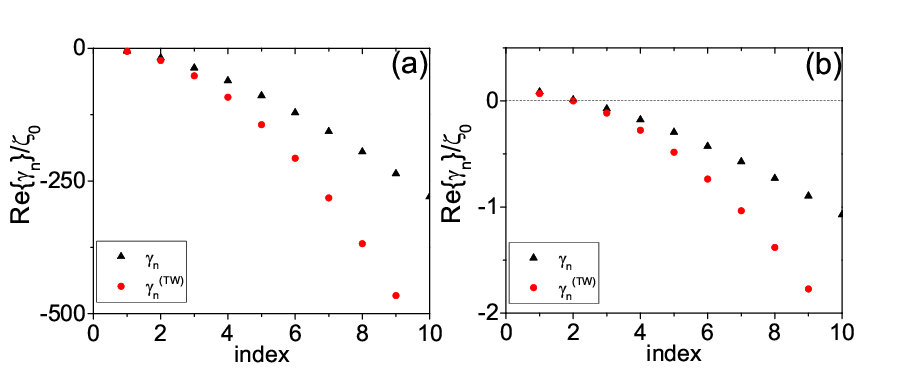}
\caption{Real part of the first 10 eigenvalues normalized to $\zeta_0=376.7 \, \text{ohm}$ for $l/w=50$, $l/\lambda=2\cdot 10^{-3}$ (a) and $l/\lambda=0.5$ (b) evaluated by the quasi-1D approach (blue triangles) and by the approximated approach (AA) (red circles).}
  \label{fig:Eig}
\end{figure}
We numerically solve the eigenvalue problem of Eq. (\ref{eq:EigenvalueProblem}) by using the Galerkin method with piecewise linear functions.  In the following, we refer to this method as quasi-1D approach.

First, we investigate two ribbons both featuring a length-to-width ratio  $l/w=50$. Specifically, the first one has a length much smaller than the operating wavelength $\lambda$, i.e. $l/\lambda=2\cdot 10^{-3}$, the second one has a length comparable to $\lambda$, i.e. $l/\lambda=0.5$. For both the investigated scenarios, in Fig. \ref{fig:Eig} we plot  the real part of the first $10$ eigenvalues $\gamma_n$ of the operator $\mathcal{F}$. We compare them with the approximated eigenvalues $\gamma_n^{\left( \text{A} \right)}$ given by Eq. (\ref{eq:EigAnl}). We note a good agreement for low index eigenvalues between the two approaches. As we increment the index $n$, the deviation between $\gamma_n$ and  $\gamma_n^{\left( \text{A} \right)}$ sensibly increases.

The first 8 eigenvalues for both scenarios are listed in Tab. \ref {tab:Eig}. For $l/\lambda = 2 \cdot 10^{-3}$, the imaginary part of the eigenvalues is much smaller in magnitude than the real part and the real part of the eigenvalues is always negative. The corresponding modes have a  quasi-electrostatic character, consistently with the fact that $l/\lambda \ll 1$. Moreover, the imaginary part of the even modes is much higher than the one of the odd modes, because odd modes have zero total dipole moment and therefore exhibit less radiation losses.  For $l/\lambda = 0.5$, the first two eigenvalues have positive real part, and their imaginary part is comparable to the real part. This fact indicates a more complex interplay between the electric and magnetic interactions (electromagnetic modes). 

\begin{table}
\caption{First 8 eigenvalues normalized to $\zeta_0=376.7 \, \text{ohm}$ for $l/w=50$, $l/\lambda=2\cdot 10^{-3}$ and $l/\lambda=0.5$.}
\begin{tabular}{c|rl|rl} 
$\gamma_n / \zeta_0$ & \multicolumn{2}{c}{$ l/\lambda = 2 \cdot 10^{-3}$}   & \multicolumn{2}{c}{$ l/\lambda 
= 0.5 $} \\
\hline
0 & $ -4.97 $&$ -5.32 \cdot 10^{-7}  i$ & $ 5.75 \cdot 10^{-2} $&$ -1.98 \cdot 10^{-2} i$  \\
1 & $ -17.9 $&$ -5.48 \cdot 10^{-12} i$ & $ 1.04 \cdot 10^{-2} $&$ -1.54 \cdot 10^{-2} i$ \\
2 & $ -36.9 $&$ -5.21 \cdot 10^{-8}  i$ & $-7.86 \cdot 10^{-2} $&$ -1.37 \cdot 10^{-6} i$   \\
3 & $ -60.9 $&$ -1.23 \cdot 10^{-12} i$ & $-1.84 \cdot 10^{-1}$&$ - 1.98 \cdot 10^{-3} i$  \\
4 & $ -89.2 $&$ -1.79 \cdot 10^{-8}  i$ & $-3.02 \cdot 10^{-1}$&$ - 2.09 \cdot 10^{-6} i$ \\
5 & $  -121 $&$ -5.37 \cdot 10^{-13} i$ & $-4.35 \cdot 10^{-1}$&$ - 7.86 \cdot 10^{-4} i$ \\
6 & $  -156 $&$ -8.84 \cdot 10^{-9}  i$ & $-5.80 \cdot 10^{-1}$&$ - 9.13 \cdot 10^{-7} i$  \\
7 & $  -195 $&$ -2.86 \cdot 10^{-13} i$ & $-7.37 \cdot 10^{-1}$&$ - 4.20 \cdot 10^{-4} i$
\end{tabular}
\label{tab:Eig}
\end{table}

\begin{figure}
\centering
\includegraphics[width=\columnwidth]{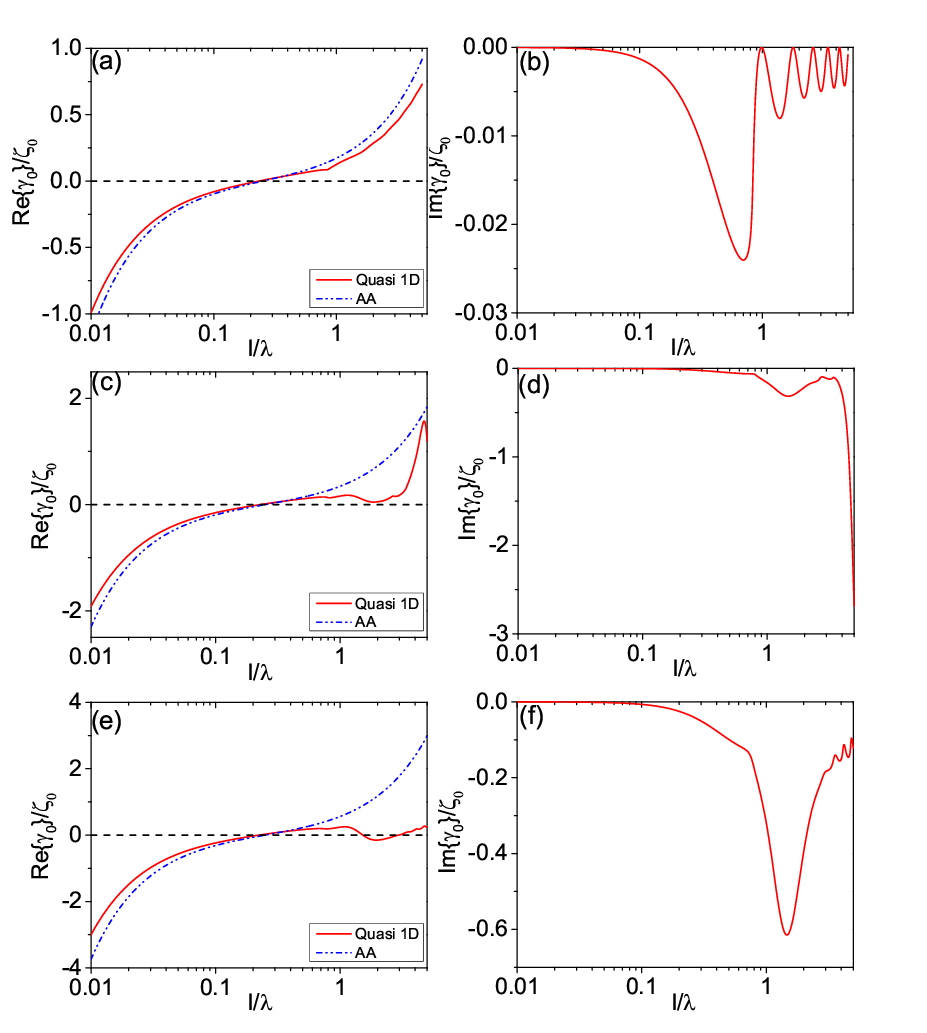}
\caption{Real and imaginary part of the first eigenvalue $\gamma_0$ normalized to $\zeta_0=376.7$ for $l/w=50$ (a),(b), $l/w=20$ (c),(d), and $l/w=10$ (e),(f) as a function of the  $l/\lambda \in \left[ 0.01, 5 \right]$. The eigenvalue has been calculated by using the quasi-1D approach (red line) and the approximated approach (AA) (blue line).}
  \label{fig:LambdaSweep}
\end{figure}

We now investigate, with the help of Fig. \ref{fig:LambdaSweep}, the dependence of the real and imaginary parts of the first eigenvalue $\gamma_0$ on $l/\lambda$ for $l/w=50,20,10$. In particular, we calculate $\gamma_0$ by considering the quasi-1D approach and the approximated approach. It is interesting to note that, by increasing $l/\lambda$, the real part of the eigenvalue $\gamma_0$ changes its sign, and becomes positive. When the real part of the eigenvalue $\gamma_0$ is zero, which approximatively occurs at $l/\lambda\approx 1/4$ according to Eq. (\ref{eq:EigAnl}), the electric energy and the magnetic energies are equal. For $l/w=10$, the real part of $\gamma_0$ is negative also when $l/\lambda \approx 1$,  in the same interval the imaginary part of $\gamma_0$ is very high. The imaginary part of the eigenvalue, which takes into account the radiation losses of the mode, is negligible for small length-to-wavelength ratios $l/\lambda$. Then, for $l/\lambda$ of the order of one, it increases and starts to oscillate.  It has a maximum magnitude that depends on $\l/w$.  High order eigevalues, not shown here, have a similar behaviour. 

Despite an overall good agreement between the eigenmodes computed numerically and those given by the approximated approach, the approximated approach is not able to correctly describe the behaviour of the charge density associated to the modes near the ends of the ribbon. Either for $l/\lambda=2 \cdot 10^{-2}$ and $l/\lambda=0.5$, in proximity of the two ends of the ribbon, the currents $u_n$ goes to zero as ${u_n^{\left(\text{REG}\right)} \left( z \right)} {\sqrt{1 - \left(z/l\right)^2}}$, while the charge densities diverge as ${q_n^{\left(\text{REG}\right)} \left( z \right)}/{\sqrt{1 - \left(z/l\right)^2}}$, where $u_n^{\left(\text{REG}\right)}$ and $q_n^{\left(\text{REG}\right)}$ are regular functions of $z$ \cite{VanBladel,Durand}. Nevertheless, the total electric charge accumulated along half-ribbon, i.e. $\int_{0}^{\pm l} q_n \left(z\right) \, dz$, remains finite. This behaviour is a structural property of the electromagnetic problem independently of the material. Furthermore, for $l / \lambda = 0.5$ the modes have a significant imaginary part that the approximated approach cannot predict.

\section{Scattering from a ribbon with tunable conductivity}
\label{sec:Virtual}
\begin{figure*}[ht!]
\centering
\includegraphics[width=0.75\textwidth]{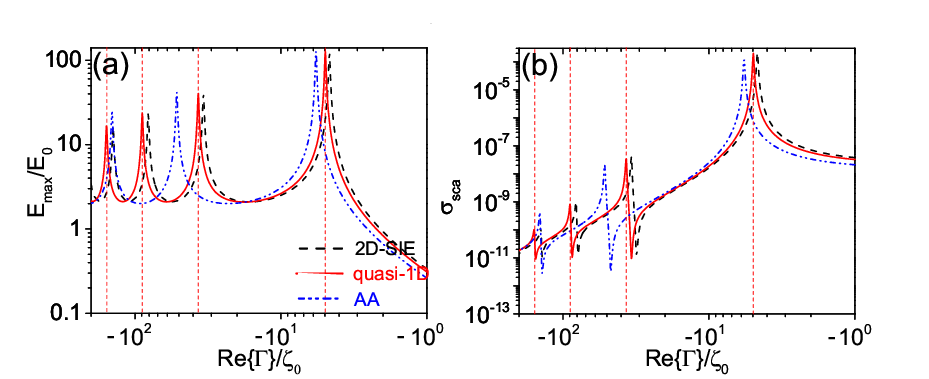}
\caption{Maximum of the magnitude of the total electric field $E_{max}$ on the ribbon (a) and scattering efficiency $\sigma_{sca}$ (b), as a function of $\re{\Gamma} \in [-2 \cdot 10^2, -1] \,  \zeta_0$ for $l/\lambda=2\cdot 10^{-3}$, $l/w=50$ and $\mbox{Im} \left\{\Gamma \right\} = 10^{-2} \mbox{Re} \left\{ \Gamma \right\}$. The two quantities are  evaluated by the quasi-1D approach (red line), by the approximated approach (AA) (blue line), and by the 2D SIE method (black line).  The plots are in loglog scale. We show with vertical dashed red lines the positions of the real part of the first four even eigenvalues $\gamma_n$, $n \in \left\{ 0, 2, 4, 6 \right\}$, ordered from the right to the left, whose values are listed in Tab. \ref{tab:Eig}.}
  \label{fig:Sweep_Chi}
\end{figure*}

\begin{figure*}[ht!]
\centering
\includegraphics[width=0.75\textwidth]{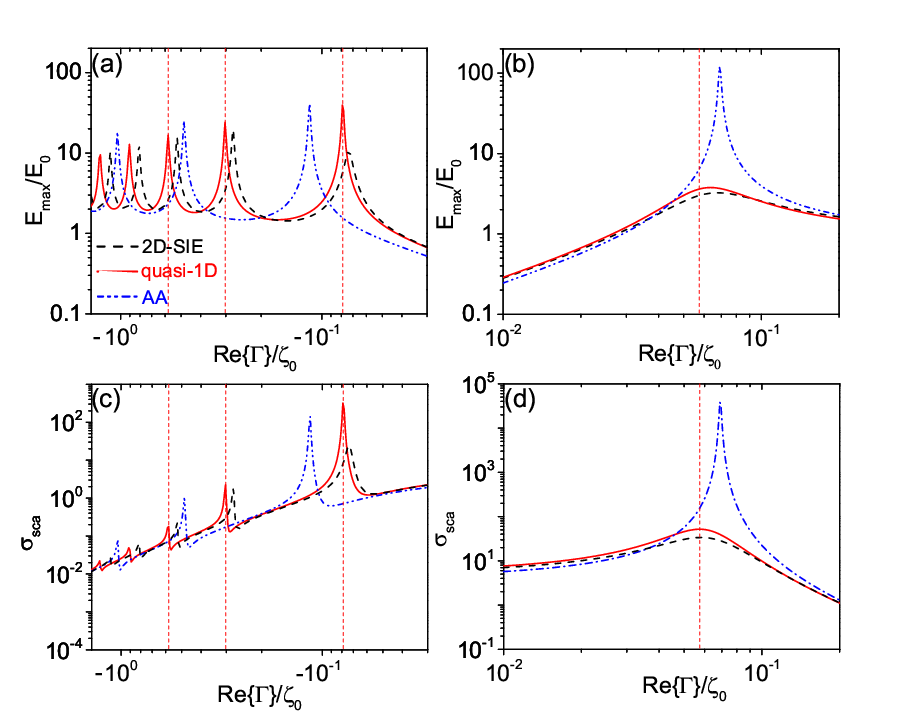}
\caption{Maximum of the magnitude of the total electric field $E_{max}$ on the ribbon (a),(b) and scattering efficiency $\sigma_{sca}$ (c),(d) as a function of $\mbox{Re} \left\{  \Gamma \right\}$ in the two intervals  $\re{\Gamma} \in \left[ -2, -0.03 \right]  \,  \zeta_0$ and $ \re{\Gamma} \in \left[ 0.01, 0.2 \right]  \,  \zeta_0$ for $l/w=50$, $l/\lambda=0.5$, and $\mbox{Im} \left\{\Gamma \right\} = 10^{-2} \mbox{Re} \left\{ \Gamma \right\}$. The two quantities are  evaluated by the quasi-1D approach (red line), by the approximated approach (AA) (blue line), and by the 2D SIE method (black line).  The plots are in loglog scale. We show with vertical dashed red lines the positions of the real part of the first four even eigenvalues $\gamma_n$, $n \in \left\{ 0, 2, 4, 6 \right\}$, whose values are listed in Tab. \ref{tab:Eig}.}
  \label{fig:Sweep_Chi2}
\end{figure*}

Now, in order to illustrate the material picture, we consider the scattering problem from a ribbon of tunable effective surface conductivity (and hence $\Gamma$) when it is  excited by a plane wave of unit intensity, polarized along the ribbon axis and propagating orthogonally to the ribbon surface. In particular, we evaluate the maximum of the magnitude of the total electric field sampled on the ribbon surface, denoted as $E_{max}$
\begin{equation}
E_{max} = \underset{ \left| z \right| < l}{\mbox{Maximum}} \left| E \left( z \right) \right|,
 \end{equation} 
and the scattering efficiency $\sigma_{sca}$,  as the parameter $\Gamma$ varies. The scattering efficiency is defined as \cite{bohren08,WriedtBook}
\begin{equation}
   \sigma_{sca} = \frac{C_{sca}}{G},
   \label{eq:ScattEffDef}
\end{equation}
where $G = 4 l w$ is the ribbon cross-sectional area, $C_{sca}$ is the scattering cross section
\begin{equation}
   C_{sca} = \frac{1}{\left\| {\bf E}_i \right\|^2} \frac{c}{\omega} \varoiint_{S_c} \hat{\bf e}_r \cdot \mbox{Im} \left\{    \left(  \boldsymbol{\nabla} \times {\bf E}_S \right)^* \times {\bf E}_S  \right\} \, \mbox{dS},
   \label{eq:ScattCrossSec}
\end{equation}
${\bf E}_S$ is the scattered field, $\hat{\bf e}_r$ is the  radial versor of a spherical reference system, and $S_c$ is an auxiliary surface enclosing the ribbon.

In Fig. \ref{fig:Sweep_Chi} we plot $E_{max}$ and $\sigma_{sca}$ for $l/\lambda = 2 \cdot 10^{-3}$ and $l/w = 50$. We restricted $\re{\Gamma}$ to vary in the interval  $\re{\Gamma} \in [-2 \cdot 10^2 \zeta_0, -\zeta_0] \,  $, chosen such that the first four even eigenmodes are resonantly excited. We assumed $\mbox{Im} \left\{\Gamma \right\} = 10^{-2} \mbox{Re} \left\{ \Gamma \right\}$.  Only the even modes with $n=0,2,4,\ldots$ are excited, the odd modes are transparent to the uniform excitation since $\langle  u_n^*, E_i \rangle = 0 \quad$ for $ n = 1,3,5, \ldots$. In this case, the inclusion of the substrate would only rescale the $\Gamma$-axis of Fig. \ref{fig:Sweep_Chi} by a factor $2/\left(1+\varepsilon_S\right)$ \cite{Jackson}, but it does not affect the values of the electric field and of the scattering efficiency.

Similarly, for the case $l/\lambda = 0.5$, we plot $E_{max}$ for $\re{\Gamma} \in \left[ -2 \zeta_0, -0.03 \zeta_0 \right]$ in Fig. \ref{fig:Sweep_Chi2} (a), and $ \re{\Gamma} \in \left[ 0.01 \zeta_0, 0.2 \zeta_0 \right]$ in Fig. \ref{fig:Sweep_Chi2} (b). In Fig. \ref{fig:Sweep_Chi2} (c),(d) we plot the corresponding  $\sigma_{sca}$.  The first even eigenmode is excited for $ \re{\Gamma} \in \left[ 0.01 \zeta_0, 0.2  \zeta_0 \right] $, and the next five even eigenmodes are excited for $\re{\Gamma} \in \left[ -2 \zeta_0, -0.03 \zeta_0 \right]$. In both cases, we show with vertical dashed red lines the positions of the real part of the first four even eigenvalues $\gamma_n$, $n \in \left\{ 0, 2, 4, 6 \right\}$, which are listed in Tab. \ref{tab:Eig}. 	

We have computed the current distribution by Eq. (\ref{eq:ModeExpansion}) (red line), by Eq. (\ref{eq:TWAnonhomSD}) (blue line) under the approximated approach, and by a 2D full-wave Surface Integral Equation (2D-SIE) method \cite{miano2005surface} (black line). The electric field on the ribbon surface is evaluated by Eq. (\ref{eq:SurfCond}).    For the scenarios presented in both Figs. (\ref{fig:Sweep_Chi}) and (\ref{fig:Sweep_Chi2}), we find very good agreement for both $E_{max}$ and $\sigma_{sca}$ between the solution of Eq. (\ref{eq:IntegroDifferential}) and the 2D-SIE approach. This fact validates our method and the corresponding numerical algorithm.  
 In Fig. \ref{fig:Sweep_Chi} (a),(b) and in Fig. \ref{fig:Sweep_Chi2} (a),(c)  the orders of magnitude of both $\sigma_{sca}$ and $E_{max}$ are correctly predicted by the approximated approach, because the material losses dominates over the radiation ones. On the other hand in Fig. \ref{fig:Sweep_Chi2} (b),(d) the approximated approach overestimates both $\sigma_{sca}$ and $E_{max}$ since, in this case, the radiation losses are dominant and are not included in the approximated approach. In all cases, the approximated approach overestimates $\re{\gamma_n}$, causing a downward shift of the peaks.

It is worth to note that, for $l/\lambda = 2 \cdot 10^{-3}$, the $\sigma_{sca}$ spectrum features asymmetric lineshapes arising from the interference of two adjacent even modes \cite{forestiere2013theory}, as shown in Fig. $\ref{fig:Sweep_Chi}$ (b). For instance,  the first dip from the right is due to the 
interference between the modes $u_0$ and $u_2$. This interference causes a cancellation of the total dipole moment of the ribbon and, therefore, a vanishing scattering because the ribbon is small compared to the incident wavelength. The lineshapes of Fig. $\ref{fig:Sweep_Chi2}$ (c) are remarkably less asymmetric with respect to Fig. $\ref{fig:Sweep_Chi}$ (b) since the ribbon is now comparable to the operating wavelength, and a cancellation of the total dipole moment do not imply zero scattering.

\section{Design Quasi-1D resonators}
\label{sec:Design}

So far, we did not make any assumption on the material composition of the ribbon, and the presented results hold for any homogeneous 2D material. We found that depending on $l/\lambda$, a given mode can be resonantly excited in materials with either negative or positive real part of $\Gamma$. In the following we consider one example of material for each scenario. In order to excite narrow resonances we consider a regime where the losses play a minor role. Therefore, we consider a frequency range where the interband transitions of the materials are negligible. 

\subsection{Graphene Layer}
In order to understand the practical implications of the introduced framework, we now consider a charge density tunable graphene ribbon \cite{vakil2011transformation} with a large number of unit cells along its the transverse direction. We assume $\mu/K_B T_0 \gg 1$ (i.e. highly gated or doped graphene), where  $\mu$ is the chemical potential, $K_B$ is the Boltzmann constant, and $T_0$ is the temperature, and we disregard the spatial dispersion.

When the effects of the intraband transition are negligible the surface conductivity of graphene takes the Drude-like form \cite{mikhailov2007new}
\begin{equation}
\sigma \left( \omega \right) \approx \frac{1}{R_0} \frac{1}{  i \Omega + \hbar/\left( \mu   \tau\right)},
\label{eq:Intraband}
\end{equation}
where $\tau$ is the electron relaxation time due to the scattering with the phonons ($\tau \approx 5 \times 10^{-13} s$), $R_0 = \pi \hbar / e^2 \cong	12.9 \, k_0 \text{ohm} $, and 
\begin{equation}
\Omega = \frac{\hbar \omega}{\mu}
\label{eq:Omega}
\end{equation}
is the normalized frequency.
The contribution of the interband term is negligible for $\Omega < 2 $. 

\begin{figure}
\centering
\includegraphics[width=0.75\columnwidth]{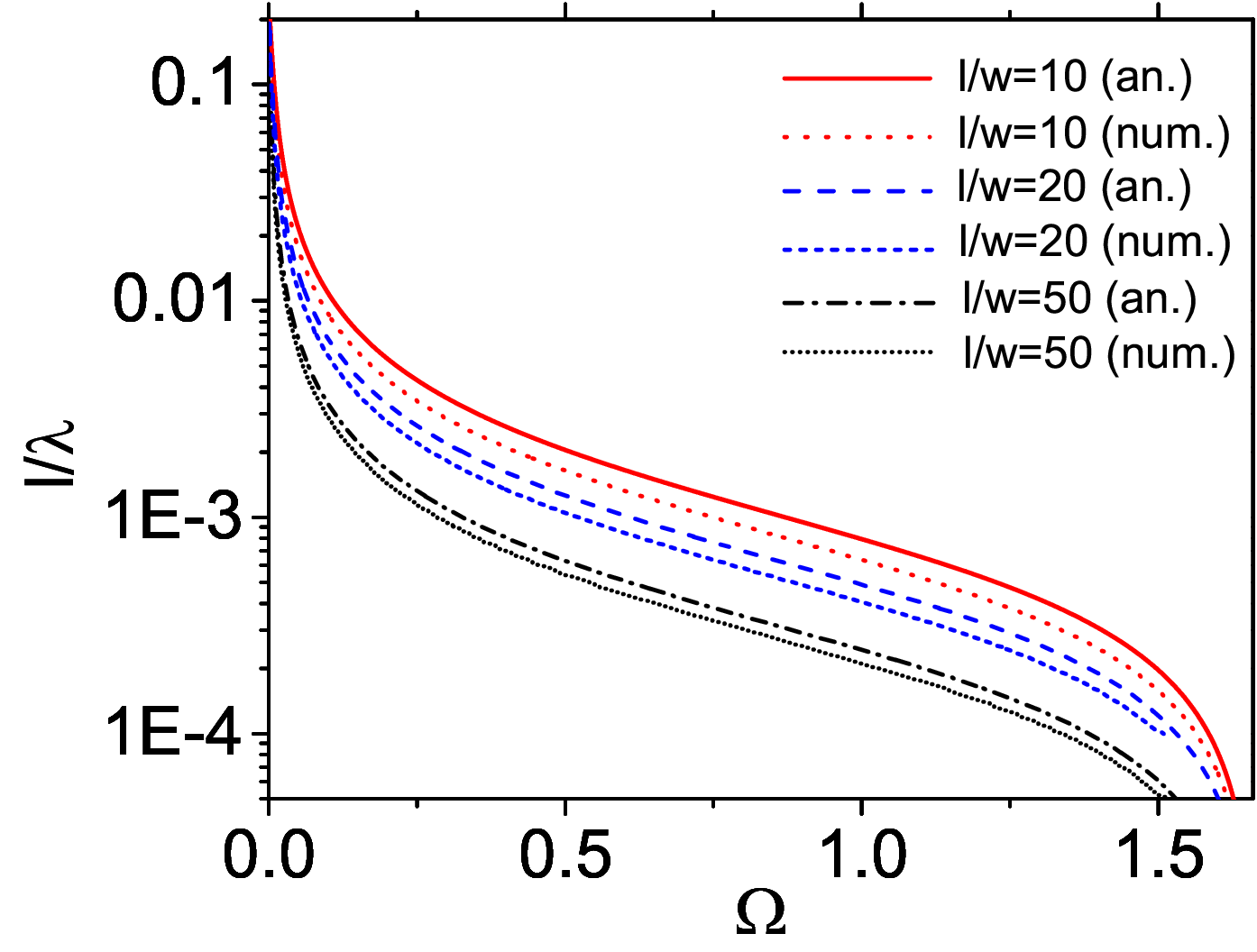}
\caption{Value of $l/\lambda$ that satisfy the resonance condition of the mode $n=0$ as $\Omega=\hbar \omega/\mu$ varies. Three different values of $l/w$ have been considered, namely $l/w=50$ (black lines), $l/w=20$ (blue lines), and $l/w=10$ (red lines). We compared the solutions of Eq. (\ref{eq:SolQuadratic}) obtained using the numerical value of $\gamma_0$ (dashed lines), with the approximated approach approximated solution given by (\ref{eq:SolQuadratic}) (continuous lines).}
  \label{fig:ResonanceGraphene}
\end{figure}

We now study the resonance conditions for the eigenmodes of the graphene ribbon when $\omega \gg 1 / \tau$.  By using Eq. (\ref{eq:Intraband}), the expression of $\Gamma$ becomes
\begin{equation}
   \Gamma = \frac{i}{\sigma} \approx - {\Omega R_0}.
   \label{eq:GammaGraphene}
\end{equation}
and the resonance condition (\ref{eq:ResonanceCondition}) ($\Gamma \approx \re{ \gamma_n }$) gives
\begin{equation}
 \Omega = -   \frac{1}{ R_0} \re{\gamma_n}.
 \label{eq:OmegaRes}
\end{equation}
Since $\re{\Gamma}<0$ only modes with $\re{ \gamma_n } <0$ can be resonantly excited. 
Under the approximated approach, Eq. (\ref{eq:OmegaRes})  becomes
\begin{equation}
\Omega = \frac{1}{A} \frac{\lambda}{l} \left[  \left( \frac{1+n}{4}\right)^2 - \left(\frac{l}{\lambda} \right)^2 \right]
\label{eq:SolQuadratic}
\end{equation}
where 
\begin{equation}
  A \left( l / w \right) =  \frac{R_0}{\zeta_0} \frac{l}{w} \frac{1}{\Theta} \approx 8.56 \frac{2l/w}{\ln\left(2l/w \right)}.
\end{equation}
Figure \ref{fig:ResonanceGraphene} shows the curves relating the values of $l/\lambda$ and $\Omega$ that satisfy the resonance condition (\ref{eq:OmegaRes}) for the $n=0$ mode and for three different values of $l/w$. We have evaluated them numerically by solving Eq. (\ref{eq:OmegaRes}) and using the numerical value of $\gamma_0$, and analytically by using the Eq. (\ref{eq:SolQuadratic}), which is based on  the approximated approach. We find satisfactory agreement between the two approaches.

Therefore, the resonant chemical potential, according to the material picture, is given by:
\begin{equation}
  \mu = \hbar \omega_l  \frac{A}{ \left( \frac{n+1}{4} \frac{\lambda}{l} \right)^2 -1},
\end{equation}
and the resonant wavelength, according to the frequency picture, is given by
\begin{equation}
 \lambda = \frac{4 l}{n+1} \sqrt{1+ A \Omega_l}, 
 \label{eq:LambdaDesign}
\end{equation}
where $\Omega_l = {\hbar \omega_l}/{\mu}$ and $\omega_l = 2\pi c_0 / l$.
\begin{table}
\caption{
Values of the normalized frequencies $\Omega$ and of the corresponding chemical potentials $\mu$ of the graphene ribbon for $w=10nm$, $l=100nm$ and $\lambda = 100 \mu m, 50 \mu m, 10 \mu m$. They are designed to enforce the resonance of the $n=0$ mode. The corresponding values of the quality factor and of the maximum field enhancement $E_{max}/E_i$ on the axis is also shown.}
\begin{tabular}{c|c|c|c|c|c} 
$\lambda (\mu m) $ & $ \gamma_0 $ & $ \Omega $ & $ \mu 
$(meV) & $\mathcal{Q}$ & $E_{max}/E_i$  \\
\hline
$100$ & $ -30.10 $ & $0.748$ & $16.6$ & $9.4$ & $10.1$  \\
$50$  & $ -15.04 $ & $0.419$ & $59.3$ & $19$ & $22.57$  \\
$10$  &  $ -3.00$ & $0.0878$ & $1410$ &  $94$ &  $111$  \\
\end{tabular}
\label{tab:Design}
\end{table}
\begin{figure}
\centering
\includegraphics[width=0.75\columnwidth]{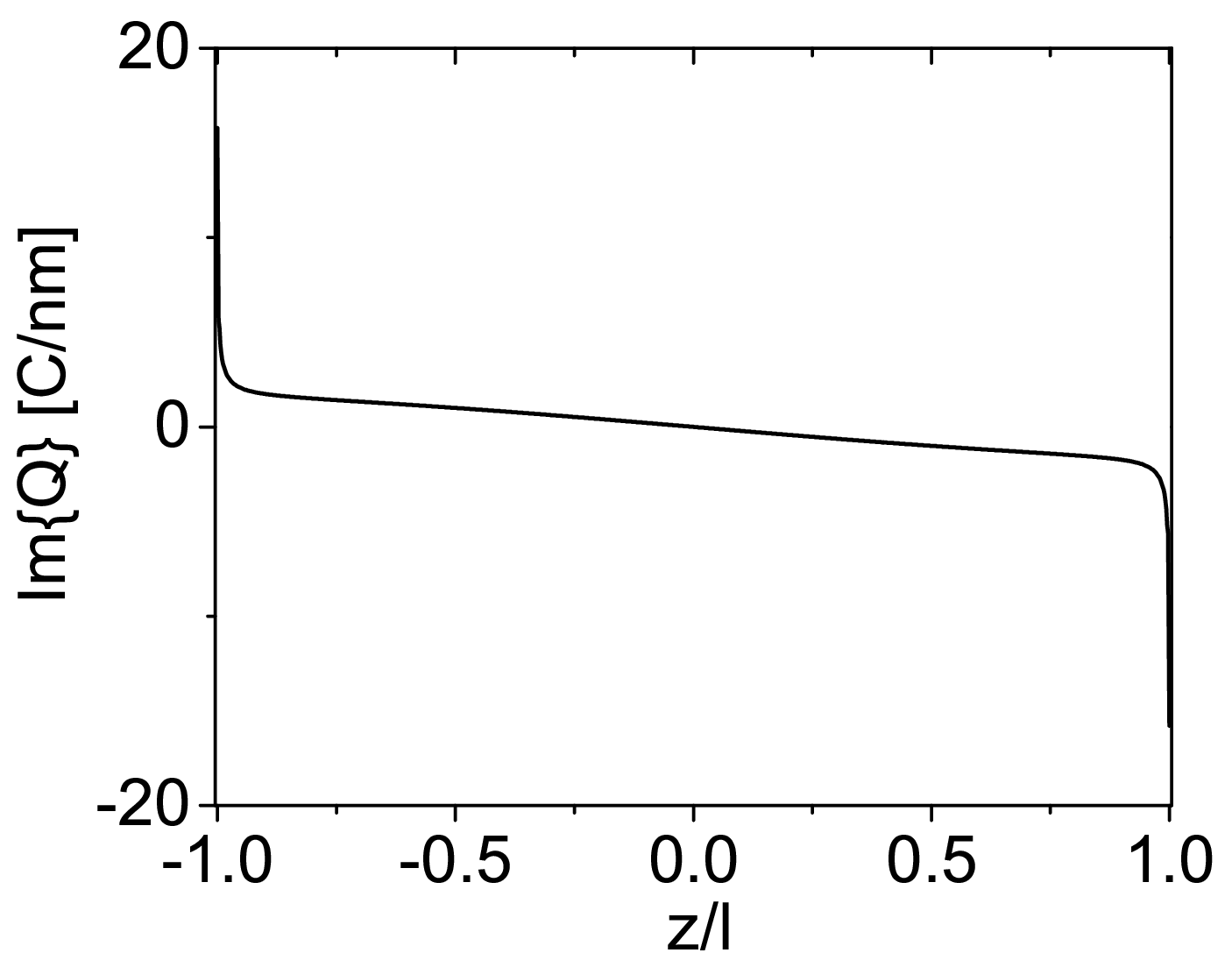}
\caption{Imaginary part of the p.u.l. charge on the graphene ribbon axis. We consider $\lambda = 50\mu m$, $\l/w=10$, $l/\lambda= 2 \cdot 10^{-3}$, $\Omega= 0.419$, and an incident z-polarized plane wave.}	
  \label{fig:GrapheneCharge}
\end{figure}
\begin{figure}
\centering
\includegraphics[width=\columnwidth]{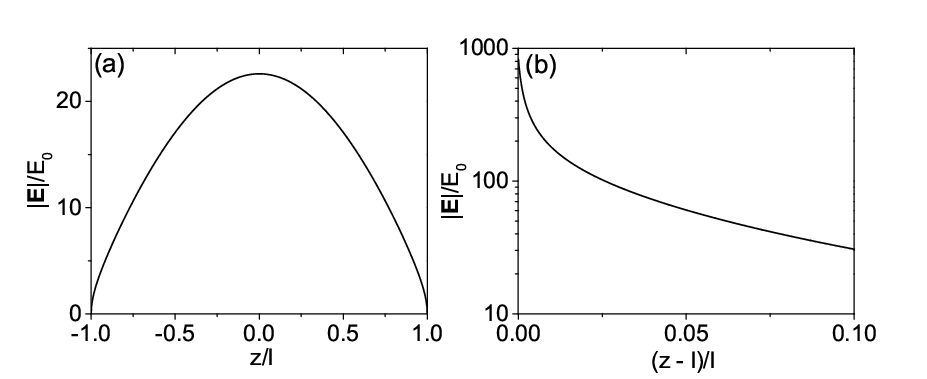}
\caption{(a) Magnitude of the total electric field normalized to $E_i$ on the graphene ribbon axis. (b) Magnitude of the electric field scattered along the horizontal red line sketched Fig. \ref{fig:Sketch} normalized to $E_i$. We consider $\lambda = 50\mu m$, $\l/w=10$, $l/\lambda= 2 \cdot 10^{-3}$, $\Omega= 0.419$ and an incident z-polarized plane wave with $E_i=1V/m$.}
  \label{fig:GraphenePlasmonic}
\end{figure}
The expression (\ref{eq:LambdaDesign}) provides the scaling laws of the graphene quasi-1D resonator in the frequency picture. In particular, for $A \Omega_l \gg 1$, the resonant wavelength scales approximately as $l$, $1/n$, $1/\sqrt{w}$, and $1/\sqrt{\mu}$.
It is interesting to compare Eq. (\ref{eq:LambdaDesign}) with the resonant condition obtained by using the dispersion relation of an infinite graphene sheet \cite{grigorenko2012graphene}

\begin{equation}
 \lambda = \left( \frac{4}{n+1} \frac{\hbar c_0 l}{\mu} \frac{R_0}{\zeta_0} \right)^{1/2}.
 \label{eq:SolQuadraticApprox2}
\end{equation}
The resonant wavelength approximatively scales as $\sqrt{l}$, $\sqrt{n}$, $1/\sqrt{\mu}$ . The different behaviour stems from electromagnetic finite size effects.

Now, we design a quasi-1D resonator based on graphene. 	First of all, we 
 consider wavelengths equal or smaller than $100 \mu m$, since in this condition the effects of the losses due to the collisions and interband transitions are negligible. We also assume $w \ge 10nm$ to neglect quantum size effects, e.g. \cite{thongrattanasiri2012quantum}. Fig. \ref{fig:ResonanceGraphene} suggests that the choice of $ l/w = 10$ corresponds to the lowest values of chemical potential. Therefore, we consider a graphene ribbon with $w=10nm$ and $l=100nm$.    In Tab. \ref{tab:Design} we list the values of the chemical potential $\mu$, which have been designed to resonantly excite the $n=0$ mode at the wavelength $\lambda = 100 \mu m, 50 \mu m, 10 \mu m$.   We also show the value of the quality factor of the resonance  $\mathcal{Q} = \omega \, \tau $, and the 
  value of the maximum electric field enhancement on the ribbon axis, $E_{max}$. The graphene ribbon is excited by a $z$-polarized plane wave with amplitude $E_i$. When the radiation losses are negligible compared to phonon scattering losses, $E_{max}$ and $\mathcal{Q}$ are closely related
\begin{equation}
 \frac{E_{max}}{E_i} \approx   \frac{4}{\pi}\frac{1}{1+n} \,  \mathcal{Q}, \qquad n =0,2,4,\ldots
\end{equation}
In particular, for the $n=0$ mode, we have $E_{max}/E_i \approx 1.3 \, \mathcal{Q}$. 

Among the three solutions listed in Tab. \ref{tab:Design}, we now investigate the scattering response of the ribbon designed to operate at $\lambda = 50 \mu m$. The designed ribbon works as a single-mode resonator. We verify this claim by comparing the 
$n=0$ term of the expansion (\ref{eq:ModeExpansion}) and the direct numerical solution of Eq. (\ref{eq:IntegroDifferential}) by using the Galerkin method. The mean square value of the difference is less than $0.04 \%$.  In Fig. \ref{fig:GrapheneCharge} we show the imaginary part of the p.u.l. charge on the ribbon axis, while its real part is negligible. The p.u.l. charge $Q$ diverges at the two ribbon ends as expected. In Fig. \ref{fig:GraphenePlasmonic} (a) we plot the magnitude of the total electric field $E(z)$ along the ribbon axis. 
In Fig. \ref{fig:GraphenePlasmonic} (b) we plot the magnitude of the scattered electric field   in the region outside the ribbon and in proximity of one of the two ends, where the charge accumulation takes place, along the red line sketched in Fig. \ref{fig:Sketch}. We show the electric field for $\left(z-l\right)/l \in \left[0 ,0.1 \right]$. 
The electric field is singular at $z=l$ because of the charge accumulation at the ribbon ends. This is the analogous of the electrostatic lightning rod effect for an edge. 

The inclusion of a silicon dioxide substrate $\left( \varepsilon_S = 3.9 \right)$ scales the eigenvalue $\gamma_0$ of the investigated free-standing graphene ribbon by a multiplicative factor $2/\left(1+\varepsilon_S\right)= 0.41$ \cite{Jackson}. Therefore, the design returns $\Omega=0.177$ and $\mu=140meV$. The resulting value of  $E_{max}/E_i$ becomes $23.34$.  The charge accumulation and the electric field singularity are not affected by the presence of the substrate.

%The model developed so far is valid for a free-standing ribbon. Nevertheless, in the limit $kl \ll 1$, it can be easily extended to a ribbon lying on a semi-infinite substrate of permittivity $\varepsilon_S$ by multiplying the static Green function $g_S$ of Eq. \ref{eq:StaticGreen} by the factor  $2/\left(1+\varepsilon_S\right)$. This result follows from the method of images for dielectrics . Therefore, in the limit $kl \ll 1$, the modes of a ribbon lying on a substrate are identical to the corresponding modes of a free standing ribbon, while the eigenvalues are the ones of a free-standing ribbon scaled of a factor  $2/\left(1+\varepsilon_S\right)$.

\subsection{Thin silicon film}

We now consider a dielectric film with thickness $t$ much smaller than the width $w$, the length $l$, and the wavelength $\lambda$. The film has a homogeneous relative permittivity $\varepsilon_r$. Under this hypothesis it is possible to model the film by a ribbon of effective surface conductivity \cite{Harrington}:
\begin{equation}
   \sigma = \frac{i k_0 t \left( \varepsilon_r - 1 \right)}{\zeta_0}.
\end{equation} 
Therefore we have:
\begin{figure}
\centering
\includegraphics[width=0.75\columnwidth]{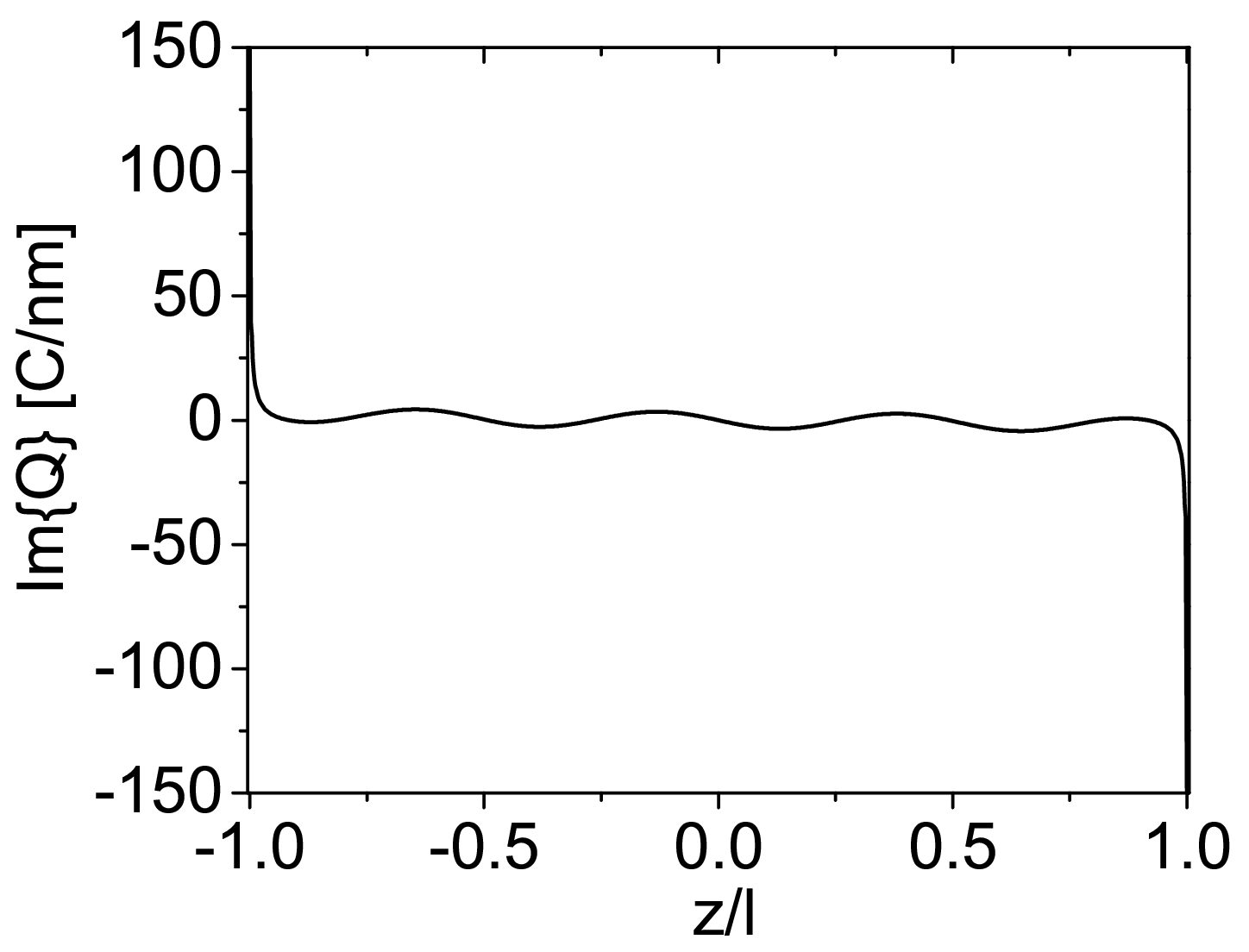}
\caption{Imaginary part of the p.u.l. polarization charge on the silicon ribbon axis. We consider $\lambda = 1\mu m$, $\l/w=10$, $l/\lambda= 3$, $t = 11.2nm$ and an incident z-polarized plane wave.
}
  \label{fig:SiliconCharge}
\end{figure}
\begin{figure}
\centering
\includegraphics[width=\columnwidth]{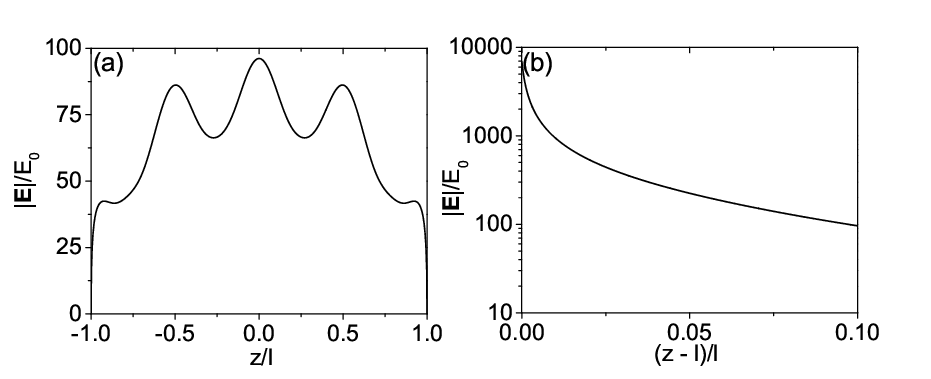}
\caption{(a) Magnitude of the total electric field normalized to $E_i$ on the silicon ribbon axis. (b) Magnitude of the electric field scattered along the horizontal red line sketched Fig. \ref{fig:Sketch} normalized to $E_i$. We consider $\lambda = 3\mu m$, $\l/w=10$, $l/\lambda= 3$, $t = 11.2nm$ and an incident z-polarized plane wave with $E_i=1V/m$.}
  \label{fig:SiliconField}
\end{figure}
\begin{equation}
  \Gamma = \frac{\zeta_0}{k_0 t \left( \varepsilon_r - 1 \right) }
  \label{eq:DielectricGamma}
\end{equation}

By considering $\lambda= 1\mu m$,  $l = 3\mu m$, and $w = 300 nm$, we have that the first bright resonant mode has $\gamma_0 =  1.2321 - 0.7787i$. At $\lambda= 1\mu m$ we have $\varepsilon_r =   12.7806 - 0.0035i $ \cite{Schinke}, thus by using Eq. (\ref{eq:DielectricGamma}) we can tune the thickness $t$ of the bar to resonantly excite the first bright mode, i.e.  $\re{\Gamma} = \re{\gamma_0}$. From the design, we obtain $t=11.2 nm$.
In Fig. \ref{fig:SiliconCharge} we plot the imaginary part of the p.u.l. polarization charge on the axis of the silicon ribbon. The polarization charge $Q$ diverges at the two ribbon ends as expected. In Fig. \ref{fig:SiliconField} (a) we plot the magnitude of the total electric field $E(z)$ along the ribbon axis. 
In Fig. \ref{fig:SiliconField} (b) we plot the magnitude of the scattered electric field in the region outside the ribbon, in proximity of one of the two ends,  along the red line sketched in Fig. \ref{fig:Sketch}. We show the electric field for $\left(z-l\right)/l \in \left[0 ,0.1 \right]$. 
Also in this case, the electric field is singular at $z=l$, because of the charge accumulation at the ribbon ends.

\section{Conclusions}
We investigated the resonance conditions of finite length nanoribbons of either conducting or dielectric material in terms of the eigenvalues and the eigenmodes of a non-Hermitian operator. We investigated the dependence of the resonances on the ribbon physical parameters. In particular, for small length-to-wavelength ratios all the eigenvalues have negative real part, while by increasing this ratio the real part of low-order eigenvalues become positive. This is significant because depending on the sign of the real part of the eigenvalue $\gamma_n$, the corresponding mode can be resonantly excited either in conductive materials if $\re{\gamma_n}<0$ or in dielectric materials if $\re{\gamma_n}>0$.  Therefore, a nanoribbon resonator may be implemented using materials with either positive or negative imaginary part of their effective surface conductivity. As an example, we investigated the scattering by two narrow and long ribbons, one made of graphene at $l/\lambda \ll 1$, and the other one made of silicon at  $l/\lambda \approx 1$.  In particular, in both cases, we designed a single mode resonator working in the infrared. It shows a strong electric field enhancement  and spectral tunability. Due to the divergence of the charge density at ribbon ends the effects of the spatial dispersion have to be considered. Its inclusion in the analysis of the resonance conditions remains an open problem from the physical and the mathematical point of view.

The unique enhancement and localization properties of the introduced quasi-1D resonator are attractive for sensing and light-matter interaction applications. In particular, field enhancement at the ribbon edges can be exploited to sense low-energy vibrational or electronic excitations of nearby molecules and to boost the  non-linear response of  nearby materials. The introduced resonator may also serve as a 1D micro/nano antenna, converting a free space propagating electromagnetic field to localized energy and vice versa. Quasi-1D  resonators may pave the way to the miniaturization of the electromagnetic circuitry, including 1D modulator and switches.

\end{document}